\documentclass[12pt]{iopart}
\usepackage{iopams}
\usepackage{graphicx}

\begin{document}

\title{$1/t$ Pressure and Fermion behaviour of Water in Two Dimensions}

\author{Taekyun~Ha$^1$, Sukmin~Chung$^1$ and M~Y~Choi$^{2,3}$}
\address{$^1$ Department of Physics, POSTECH, Pohang 790-784, Republic of Korea}
\address{$^2$ Department of Physics and Astronomy, Seoul National University, Seoul 151-747, Republic of Korea}
\address{$^3$Asia Pacific Center for Theoretical Physics, Pohang 790-784, Republic of Korea}
\ead{smc@postech.ac.kr}

\begin{abstract}
A variety of metal vacuum systems displays the celebrated $1/t$
pressure, namely, power-law dependence on time $t$, with the
exponent close to unity, as to the origin of which there has been
long-standing controversy. Here we propose a chemisorption model for
water adsorbates, based on the argument for fermion behaviour of
water vapour adsorbed on a stainless-steel surface, and obtain
analytically the power-law behaviour of pressure, with an exponent
unity. Further, the model predicts that the pressure should depend
on the temperature $T$ according to $T^{3/2}$, which is indeed
confirmed by our experiment. Our results should help elucidate the
unique characteristics of the adsorbed water.
\end{abstract}


\section*{}

Water is a natural wonder; this makes its unusual behaviour
of great interest, let alone the structure of the water molecule
that helps to explain this behaviour \cite{zubavicus2004}. Unlike
other common liquids, not only does water expand and become less
dense as it cools from 4$\,^\circ$C to 0$\,^\circ$C, it becomes even
less dense as it freezes to ice \cite{cho1997}. Although the
uniqueness of water inheres largely in its liquid phase, such
uniqueness can also be found in its adsorbed phase. Water adsorption
has been observed to produce $1/t$ pressure (i.e., pressure $p$ with
a power-law dependence on time $t$: $p \sim t^{-\alpha}$ with ${\bf
\alpha}$ close to unity) in a tremendous variety of metal vacuum
systems \cite{Hudson,O'Hanlon}. As simple as the function may seem,
scientists have long been baffled about why the pressure obeys such
a power-law \cite{Dayton,Horikoshi,Li1994,Kanazawa,Redhead}.
Despite 60 years of intense effort, however, the nature of their
solution remains obscure. In this communication, we argue that water vapour,
when adsorbed on a stainless-steel surface, behaves as fermions in
two dimensions and, on this basis, propose a chemisorption model for
water adsorbates. Remarkably, there follows analytically the
power-law behaviour of the pressure, with an exponent $\alpha=1$.
Further, the model predicts that the pressure should depend on the
temperature $T$ according to $T^{3/2}$. Indeed, this prediction is
borne out by our experimental measurements.

\begin{figure}
\center\includegraphics[width=16.0cm]{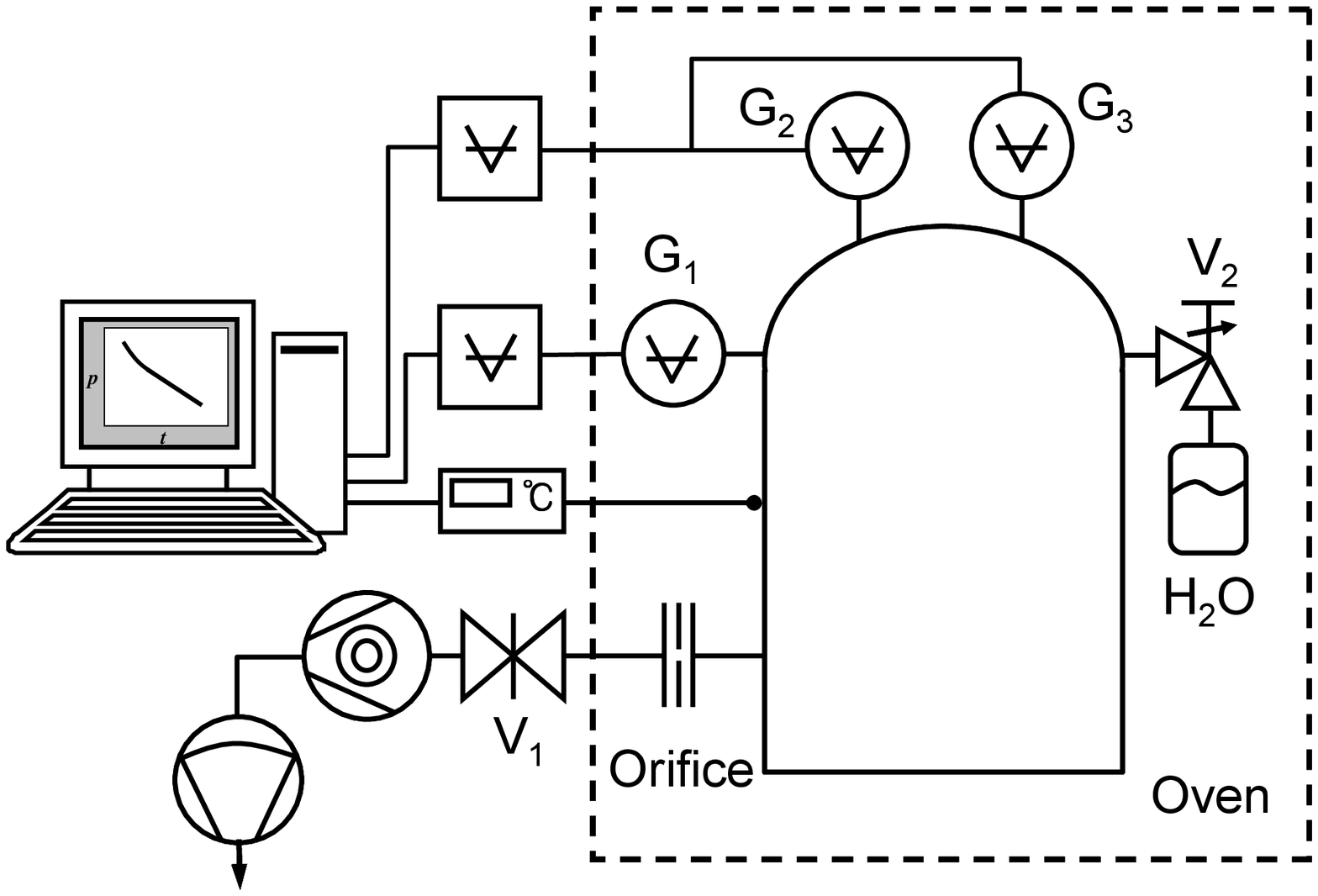}
\caption{ \label{fig1} Schematic of the metal (stainless-steel)
vacuum system used in this experiment. The test chamber has a volume
of about 42\,L and an internal surface of about 10\,600\,cm$^2$
whereas the diameter of the pumping orifice placed at the end of a
short high-vacuum line, 8\,cm long of 15\,cm diameter, is 0.8\,cm.
Pressure, which is nitrogen equivalent, is measured with the
extractor gauge (G$_1$). Measurements are also performed with two
capacitance manometer gauges (G$_2$ and G$_3$); in particular, G$_3$
is used as a transfer standard from 0.05\,Torr to $10^{-5}$\,Torr. }
\end{figure}

\begin{figure}
\center\includegraphics[width=16.0cm]{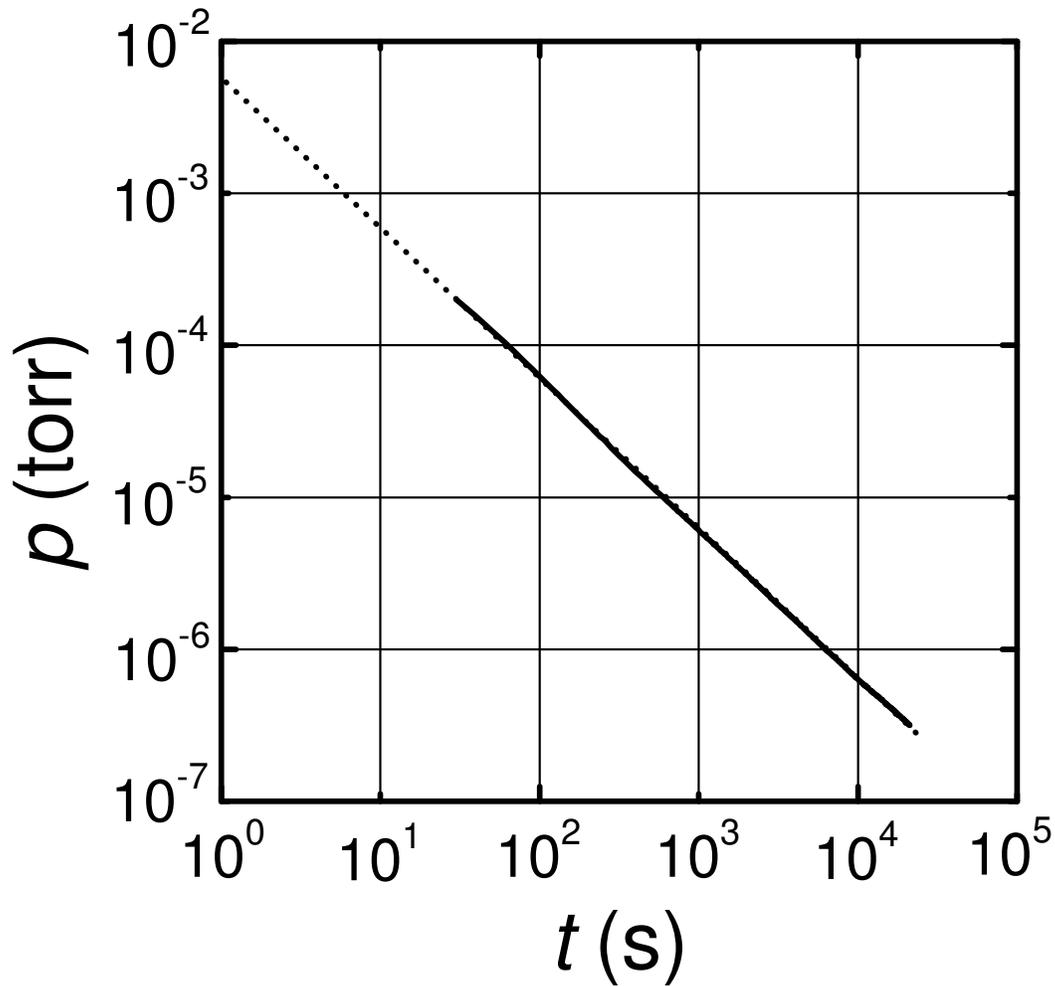}
\caption{ \label{fig2} Sample set of pump-down curves, displaying
the time evolution of the system pressure on a logarithmic scale.
The pressure follows $p(t)\sim t^{-\alpha}$ with $\alpha =1$. The
dashed line represents the extrapolation to estimate the intercept
on the $p$-axis.}
\end{figure}

The most direct examination of the
adsorption behaviour of water is to measure the pump-down
characteristics of a system. The experimental work has been
performed on an extreme high vacuum (XHV) chamber shown in
figure~\ref{fig1}, a stainless-steel vessel with a thin, dense
Cr$_2$O$_3$ film at the inner surface. This film has very few
surface singularities and a very low specific surface area
\cite{blcho1999}. Prior to each experimental run the chamber is
heated to about $100\,^\circ$C for about 6\,h and then cooled. At a
pressure below $10^{-8}$ Torr, deionized water is admitted through a
UHV leak valve to the chamber. As the adsorption proceeds, the
pressure in the chamber decreases quite slowly after the rapid
initial increase, then levels off after about 24\,h, when it
approaches its ``terminal value.'' Since the chemical potential and
hence the initial number of adsorbates must be the same for each
run, further adjustments are made to the terminal pressure. During
the run, the temperature of the vacuum system is regulated to $\pm
0.2$\,K. Figure~\ref{fig2} shows a typical pump-down curve for the
XHV system with adsorbed water. Immediately after pump-down starts,
most of the gas which goes into the pumping system is the volume
gas. Though not shown here, the pressure $p$ in the system falls
exponentially with time $t$. Subsequently, as surface desorption
takes over as the predominant source of the gas load, the fall
becomes less marked and the pressure follows the power-law $p\sim
t^{-1}$, as indicated in figure~\ref{fig2}. It is this power-law that
makes adsorbed water deserve special attention. The emerging
power-law indicates that the system of adsorption, being far from
random, organizes itself to criticality. At this stage the pressure
still decreases, but only slowly, and the system is said to be in
the quasi-steady state \cite{Kanazawa}.

The mathematical description of the desorption process begins with
the pumping equation. We suppose that the pump evacuates a vacuum
vessel of volume $V$, which has long been exposed to water, with the
effective pumping speed $S$. At any instant the system pressure $p$
is governed by the competition between the rate of gas supply due to
desorption and the rate of depletion due to adsorption on the walls
of the vessel as well as through pumping by the pump. Particle
conservation thus yields
\begin{equation}
 \label{eq1}
 V \frac{\rmd p}{\rmd t}=-\frac{\rmd N}{\rmd t} kT - sA\Gamma kT - Sp
                =-kT \frac{\rmd N}{\rmd t}- (R+S)p
\end{equation}
at temperature $T$ (with $k$ being the Boltzmann constant), where
$N$ denotes the total number of adsorbed molecules, $s$ the sticking
probability, and $A$ the inner surface area of the vacuum chamber.
According to the kinetic gas theory, the rate $\Gamma$ at which
particles, each of mass $m$, impinge on a surface (per unit area and
time) is given by $\Gamma = p/\sqrt{2\pi mkT}$, thus the
re-adsorption rate (per unit density) is given by $R \equiv
sA\sqrt{kT/2\pi m}$. Under the quasi-steady state conditions, where
the change in pressure is very slow, the pressure $p$ and the
sticking probability $s$ may be regarded as effectively constant
during the variation of $N$. Equation~(\ref{eq1}) then reduces to
\begin{equation}
 \label{eq2}
 p = -\frac{kT}{R+S} \frac{\rmd N}{\rmd t}.
\end{equation}
Thus, studying the desorption rate directly yields interesting
information about the pressure and vice versa.

We need to assess the desorption rate. For simplicity, no gas is
assumed to diffuse out of the bulk of the system walls and
adsorbates are considered confined to the (two-dimensional) surface
of the wall. Unfortunately, the details as to how water molecules
are adsorbed on a stainless-steel surface are not known: Whereas
according to the traditional view, a few, say, $\nu$ water molecules
form a cluster via hydrogen bonding, it is not clear whether some of
the $\nu$ molecules should dissociate into hydrogen atoms (H) and
hydroxyl groups (OH) so that local bonding could occur with
individual fragments \cite{Menzel2002,Feibelman2002,Materzanini2005}.
In any case, some oxygen atoms are expected to
bind to the surface atoms through hybridization. It then appears
reasonable to assume that only one particle (i.e., fragment of a few
water molecules and possibly hydroxyl groups) can occupy each
location of the surface, binding through hybridization. The binding
energy for an additional particle is expected to be substantially
smaller, which means that rather higher energy is involved for
additional occupancy; disregarding this results in the
single-occupancy condition, which forbids multi-layer adsorption and
may be taken effectively into account by introducing a hard core to
each particle. Accordingly, upon adsorption, each (fragment)
particle, in general consisting of an even number of fermions
(regardless of the presence/absence of dissociation), may be
regarded as a boson with a de facto hard core taking care of the
single occupancy condition. The total number of those bosons is
given by $\tilde{N} = N/\nu$; hereafter the tilde sign will be
omitted for simplicity.

Note that in two dimensions such a boson with a hard core is
equivalent to a fermion in an appropriate gauge field, as follows
\cite{Fradkin,Choi,Odintsov}: In the second quantized
representation the system is described by the boson operators
$b^{\dagger}_i$ and $b_i$ creating and annihilating a boson at
(surface) site $i$, respectively. The single occupancy condition
reads $n_i \equiv b^{\dagger}_i b_i = 0, 1$ or $(b^{\dagger}_i)^2 =
b_i^2 = 0$. The Jordan-Wigner transformation then maps such a boson
system into a fermion system in the gauge field corresponding to the
flux per plaquette given by $\Phi =(\alpha/\gamma_c )\sum_i n_i$ in
units of the flux quantum (i.e., $\Phi_0 \equiv 1$), where $\alpha$
is an odd integer, $\gamma_c$ denotes the coordination number, and
the summation is over $\gamma_c$ sites around the plaquette.
The interactions between adsorbate (fragment) particles, which are
neutral, are expected to be weak and negligible compared with the
interactions with the surface. Neglecting the latter as well gives
two-dimensional (2D) free fermions. Conversely, in the limit of
strong interactions with the surface, the system reduces to 2D
tight-binding fermions. Since the two opposite limiting cases give
mostly the same results, we may take the system to be just free
fermions for simplicity.\footnote{To be precise, the interplay of the surface potential and
the gauge field leads the system to be described by Harper's
equation in both the two limits, with the frustration parameters
reciprocal to each other. As a result, some of the degeneracy of
each level is lifted, if the flux $\Phi$ is not an integer, yielding
sublevels. This, however, does not alter the subsequent analysis
based on the continuum approach.}
We are thus left with 2D free fermions in a gauge field, which are
known to form Landau levels.  The mean occupation number is then
given by the Fermi function $f(\varepsilon_n)=
g[\rme^{\beta(\varepsilon_n -\mu)}+1]^{-1}$ with $\beta\equiv 1/kT$,
where $g$ denotes the degeneracy factor of each level,
$\varepsilon_n = (2n+1)\varepsilon_0$ the energy at the $n$th Landau
level, and $\mu$ the ``effective'' chemical potential of an
adsorbate particle with the binding energy $\varepsilon_{\rm b}$ included.
(Thus the ``bare'' chemical potential is given by $\mu_{\rm b} = \mu
-\varepsilon_{\rm b}$.)

According to equation~(\ref{eq2}), the pressure depends on the desorption
rate. As the vessel is evacuated by pumping, some adsorbate
particles tend to desorb, escaping dominantly via thermal activation
at room temperature. For a given energy, the number of adsorbate
particles reduces with time, proportionally to $\rme^{-t/\tau}$ with
the characteristic time $\tau$ measuring the average lifetime on the
surface. It depends on the energy of the particle:
In the case of thermal activation, the activation probability is proportional to the Boltzmann factor $\rme^{-\beta\Delta U}$, where $\Delta U$ is the energy barrier. For a particle with energy $\varepsilon_n$, it is given by $\Delta U = \varepsilon_{\rm b} -\varepsilon_n$. Accordingly, the average lifetime on the surface, which is inversely proportional to the activation probability, takes the form $\tau(\varepsilon_n) = \tau_0 \rme^{\beta \Delta U} = \tau_0 \rme^{\beta (\varepsilon_{\rm b} -\varepsilon_n)}$, where $\tau_0$ is a characteristic ``attempt time'', usually of the order of an inverse phonon frequency for many activated processes in solids.

The total number of adsorbate particles at time $t$ is then given by
\numparts
\begin{equation}
 \label{eq3a}
 N =\sum_{n=0}^{\infty} \rme^{-t/\tau(\varepsilon_n )} f(\varepsilon_n) =
 g \sum_{n=0}^{\infty}\frac{\rme^{-t/\tau(\varepsilon_n)}}{\rme^{\beta(\varepsilon_n-\mu)}+1},
\end{equation}
which, via the Euler-Maclaurin formula, is expanded as
\begin{equation}
 \label{eq3b}
\fl\qquad\qquad  N = D \int_{0}^{\infty} \rmd\varepsilon
   \frac{\rme^{-t/\tau(\varepsilon)}}{\rme^{\beta(\varepsilon -\mu)}+1}
   - \frac{\beta\varepsilon_{0}^2 D}{6}
   \frac{\rme^{-t/\tau(0)}}{1+\rme^{-\beta\mu}}
           \left(\frac{t}{\tau_0}\rme^{-\beta\varepsilon_{\rm b}}+\frac{1}{\rme^{\beta\mu}+1}
             +\cdots \right).\nonumber
\end{equation}
\endnumparts
Note that the ground-state energy $\varepsilon_0$ and the degeneracy
factor $g$ are related via $g=2\varepsilon_0 D$, where $D= \nu
mA/2\pi\hbar^2$ is the (constant) density of states in two
dimensions. Taking the derivative with respect to time gives the
desorption rate
\begin{equation}
 \label{eq4}
\fl\qquad -\frac{\rmd N}{\rmd t} = D \int_{0}^{\infty}
 \frac{\rmd\varepsilon}{\tau(\varepsilon)}
   \frac{\rme^{-t/\tau(\varepsilon)}}{\rme^{\beta(\varepsilon -\mu)}+1}
    -\frac{\beta\varepsilon_{0}^2 D}{6\tau_0}\frac{\rme^{-\beta\varepsilon_{\rm b} -t/\tau(0)}}{1+\rme^{-\beta\mu}}
        (\frac{t}{\tau_0}\rme^{-\beta\varepsilon_{\rm b}}-\frac{1}{1+\rme^{-\beta\mu}}+\cdots).\nonumber
\end{equation}
Since the binding energy is of the order of 1\,eV, the (effective)
chemical potential is much larger than the thermal energy $kT$ at
room temperature, i.e., $\beta\mu \gg 1$. Then the Fermi function in
the right-hand side of equation~(\ref{eq4}) reduces to the step function:
$f(\varepsilon) = g[\rme^{\beta(\varepsilon -\mu)}+1]^{-1} \approx
g\theta(\mu{-}\varepsilon)$. On the other hand, since the factor
$\rme^{-\beta\varepsilon_{\rm b}}\ll 1$, the second term in the right-hand
side of equation~(\ref{eq4}) is negligible compared with the first term.
Thus equation~(\ref{eq4}) becomes
\begin{equation}
 \label{eq5}
 -\frac{\rmd N}{\rmd t}\approx D \int_{0}^{\mu}
 \frac{\rmd\varepsilon}{\tau(\varepsilon)} \rme^{-t/\tau(\varepsilon)}
 = DkT \int_{\tau(\mu)}^{\tau(0)}
 \frac{\rmd\tau}{\tau^2} \rme^{-t/\tau}.
\end{equation}
Typically, we have $\tau_0 \sim 10^{-13}$\,s; thus $\tau(\mu)\equiv
\tau_{0}\rme^{-\beta(\mu-\varepsilon_{\rm b})}$ and $\tau(0) \equiv
\tau_{0}\rme^{\beta\varepsilon_{\rm b}}$ are sufficiently smaller and larger
than the observation time in figure~\ref{fig2}, respectively. The
lower and upper limits of the integral in equation~(\ref{eq5}) can then be
replaced by 0 and $\infty$ with negligible error, leading to the
desorption rate
\begin{equation}
 \label{eq6}
 -\frac{\rmd N}{\rmd t}= \frac{DkT}{t}.
\end{equation}
Substituting equation~(\ref{eq6}) into equation~(\ref{eq2}), we obtain
\begin{equation}
 \label{eq7}
 p = -\frac{\nu kT}{R+S}\frac{\rmd N}{\rmd t}
   = \frac{\nu Dk^{2}T^{2}}{(R+S)t}
  \equiv \gamma t^{-1},
\end{equation}
which is the desired $1/t$ pressure.

\begin{figure}
\center\includegraphics[width=16.0cm]{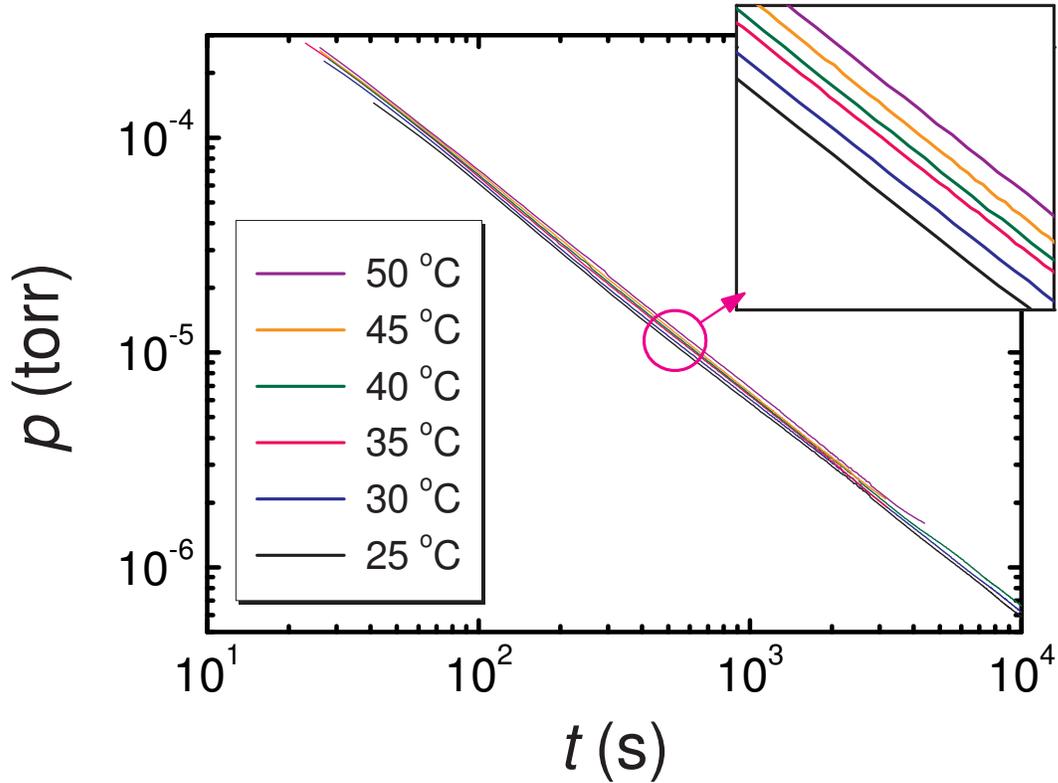}
\caption{ \label{fig3} Pump-down curves at various temperatures. The
initial number of adsorbates and hence the chemical potential
[related by equation~(\ref{eq3a}) at $t=0$] should be the same for each
experimental run. We choose the numeric value of the (bare) chemical
potential $\mu_{\rm b} =-0.677$\,eV and determine the initial pressure
according to: $p = (mkT/2\pi\hbar^2 )^{3/2} kT \rme^{\mu_{\rm b}/kT}$. Note
that pumping performance tends to degrade with temperature, as all
thermally activated processes are accelerated. The inset reveals
that this shows up as a shift to the right of the pump-down curves.}
\end{figure}

\begin{figure}
\center\includegraphics[width=16cm]{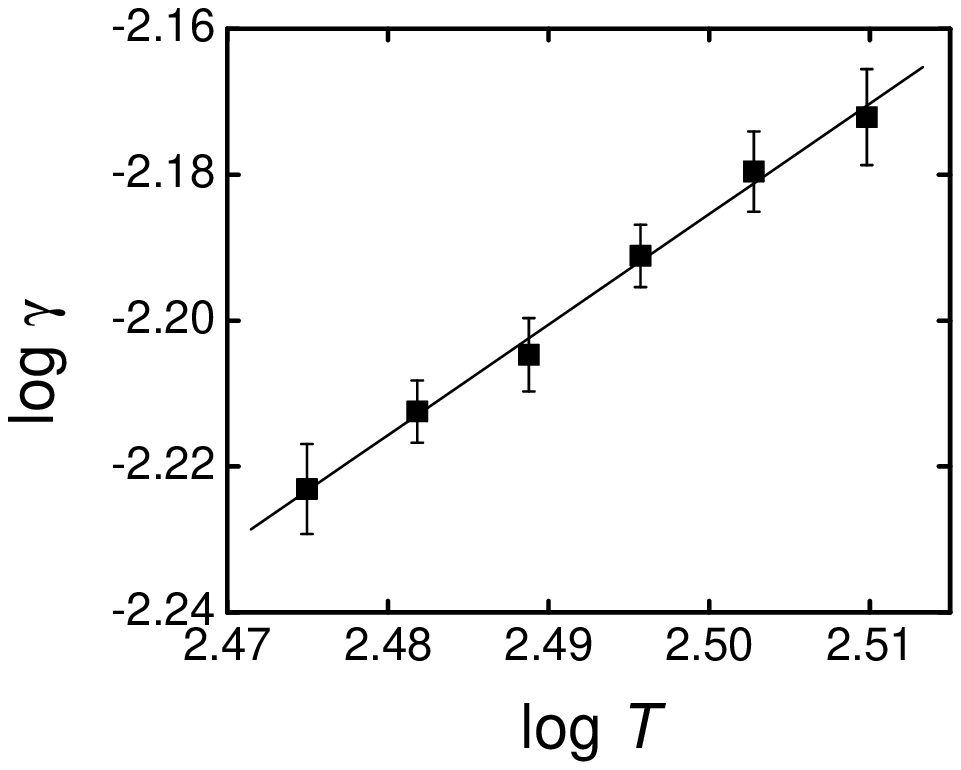}
\caption{\label{fig4} Proportionality coefficient $\gamma$ in
equation~(\ref{eq8}) versus temperature $T$, presented in the logarithmic
scale. To estimate $\gamma$, we have extrapolated the pump-down
curve, as shown by the dashed line in figure~\ref{fig2}. Each data
point consists of five independent measurements and error bar
indicate standard deviation. The solid line represents the
least-squares fit of the data, with the slope $1.515$.}
\end{figure}

We now make a quantitative test of our assumptions by probing the
temperature dependence of the proportionality coefficient $\gamma$.
This temperature dependence has been measured for the pump-down of
adsorbed water. At different temperatures, equation~(\ref{eq7}) gives a
group of parallel straight lines on the log-log plot, as shown in
figure~\ref{fig3}. The conductance and hence the pumping speed for
molecular flow of an orifice is proportional to the mean speed of
gas molecules. Namely, $S$ as well as $R$ is proportional to
$\sqrt{T}$, thus $\gamma$ takes the form
\begin{equation}
 \label{eq8}
 \gamma =\frac{\nu Dk^{2}T^{2}}{R+S} = C T^{3/2}
\end{equation}
with constant $C$ or
\begin{equation}
 \label{eq9}
 \log\gamma = \log C + \frac{3}{2} \log T,
\end{equation}
which indicates a linear dependence of $\log\gamma$ on $\log T$ with
slope $3/2$. Figure~\ref{fig4} indeed reveals a linear relation
between the two, with the slope estimated as $1.515 \pm 0.198$. Such
excellent agreement with experiment strongly supports the validity
of our analytical approach, particularly the argument for effective
fermion behaviour.

As a further check of equation~(\ref{eq7}), the adsorption isotherm is
computed and compared with the empirical isotherm in the existing
literature. Integrating equation~(\ref{eq6}) and using equation~(\ref{eq7}) to
eliminate $t$, we obtain $N$ as a function of $p$ and $T$:
\begin{equation}
 \label{eq10}
 N = DkT \ln\left(\frac{R+S}{\nu Dk^2 T^2}p\right)
\end{equation}
up to an additive constant. At a set temperature equation~(\ref{eq10})
essentially describes the Temkin isotherm between the number of
adsorbate particles and the pressure, which is known to adequately
reproduce the experimental $1/t$ behaviour over restricted pressure
ranges \cite{Kanazawa,Redhead,Kraus,Schram,Elsey}. Further, a
close look at the proportionality constant shows that the Temkin
isotherm also predicts the $T^{3/2}$ dependence. Note, however, that
the Temkin isotherm is obtained by assuming without substantiation a
constant density of sites over a wide range of energy
\cite{Redhead}. It is remarkable that such a constant distribution
in fact corresponds to the Fermi function here. In this respect, our
analytical approach also provides a theoretical basis for the
empirically-obtained isotherm.

The power-law behaviour of the $1/t$ pressure indicates that a
characteristic time scale does not exist. This can be seen by
examining the distribution $g(\tau)$ of the average surface lifetime
$\tau$, which depends on the energy $\varepsilon$. It is related
with the energy distribution, which is just the Fermi function
$f(\varepsilon)$, via $g(\tau) =
f(\varepsilon)|\rmd\tau/\rmd\varepsilon|^{-1}$. This yields
\begin{equation}
 \label{eq11}
g(\tau) \propto \tau^{-1}
\end{equation}
in the rather wide range $\tau (\mu) \lesssim \tau \lesssim \tau
(0)$. Thus the lifetime distribution in adsorption is scale free,
following a power-law with exponent unity: The number of adsorbates
with given surface lifetime is inversely proportional to the
lifetime. Such absence of a characteristic lifetime in turn gives
rise to the power-law behaviour of the pressure.

In this article, we have probed both experimentally and
theoretically the $1/t$ pressure, observed frequently in systems
with adsorbed water. Theoretical focus on a strongly bonded surface
monolayer allows us to regard adsorbates as bosons with de facto
hard cores, which transform into fermions in two dimensions. The
$1/t$ pressure as well as the power-law distribution of surface
lifetimes has then been obtained analytically, as a consequence of
the fermion behaviour of adsorbed water. Accurate measurements have
been carried out with the XHV system to confirm the validity of this
theoretical analysis. In particular, the $T^{3/2}$ dependence
observed in the measurement has been shown to be fully consistent
with the prediction of the theoretical analysis.
While the structure of water adsorbed on a well-defined metal
surface remains as a subject of debate \cite{Menzel2002,Feibelman2002,Materzanini2005}, our results suggest that water,
upon adsorption on a stainless-steel surface, exhibits fermion
behaviour. This gives rise to criticality without a characteristic
time scale, which provides an explanation of the power-law behaviour
with exponent unity. Thus, like liquid and solid water, adsorbed
water also exhibits unique characteristics.
Based on the theoretical analysis, we expect similar adsorption characteristics of other molecular systems as well: a few molecules form a small network and only one of the resulting ``particles'' can occupy a site, binding rather strongly, e.g., through hybridization.
The adsorption system controlled by the vapour of appropriate molecules should be prepared carefully in XHV, with water vapour removed thoroughly.
It would be of interest to build such a system and examine adsorption and corresponding pressure behaviours.

\begin{ack}
One of us (M.Y.C.) thanks D. Ham for hospitality during his
stay at Harvard University, where part of this work was accomplished.
This work was supported in part by the Human Resource Training
Project of Korea Industrial Technology Foundation for Regional
Innovation and in part by the BK21 Program.
\end{ack}

\clearpage

\clearpage

\clearpage

\clearpage


\section*{References}

\begin{thebibliography}{99}

\bibitem{zubavicus2004} Zubavicus~Y and Grunze~M 2004 {\it Science} {\bf 304} 974
\bibitem{cho1997} Cho~C~H, Singh~S and Robinson~G~W 1997 \JCP {\bf 107} 7979
\bibitem{Hudson} Hudson~J~B 1998 {\it Foundations of Vacuum Science and Technology} ed Lafferty~J~M (New York: John Wiley \& Sons) p~547
\bibitem{O'Hanlon} O'Hanlon~J~F 1989 {\it A User's Guide to Vacuum Technology} (New York: John Wiley \& Sons)
\bibitem{Dayton} Dayton~B~B 1961 {\it Proc. 2th. Int. Vacuum Congress} (New York: Pergamon) p~42
\bibitem{Horikoshi} Horikoshi~G 1987 {\it J. Vac. Sci. Technol.} A {\bf5} 2501
\bibitem{Li1994} Li~M and Dylla~H~F 1994 {\it J. Vac. Sci. Technol.} A {\bf12} 1772
\bibitem{Kanazawa} Kanazawa~K 1989 {\it J. Vac. Sci. Technol.} A {\bf7} 3361
\bibitem{Redhead} Redhead~P~A 1995 {\it J. Vac. Sci. Technol.} A {\bf13} 467
\bibitem{blcho1999} Cho~B \etal 1999 {\it Surf. Sci.} {\bf439} L799
\bibitem{Menzel2002} Menzel~D 2002 {\it Science} {\bf295} 58
\bibitem{Feibelman2002} Feibelman~P~J 2002 {\it Science} {\bf295} 99
\bibitem{Materzanini2005} Materzanini~G \etal 2005 {\it Phys. Rev.} B {\bf71} 155414
\bibitem{Fradkin} Fradkin~E 1989 {\it Phys. Rev. Lett.} {\bf63} 322
\bibitem{Choi} Choi~M~Y 1994 {\it Phys. Rev.} B {\bf50} 10088
\bibitem{Odintsov} Odintsov~A~A and Nazarov~Y~V 1995 {\it Phys. Rev.} B {\bf51} 1133
\bibitem{Kraus} Kraus~T 1958 {\it Transactions of the 5th National Vacuum Symposium, AVS} (New York: Pergamon) p~38
\bibitem{Schram} Schram~A 1963 The {\it Le Vide} {\bf103} 55
\bibitem{Elsey} Elsey~R~J 1975 {\it Vacuum} {\bf25} 299

\end{thebibliography}
\end{document}